\documentclass{article}

\usepackage{PRIMEarxiv}

\newcommand{\abs}[1]{\left\vert#1\right\vert}

\newcommand{\mc}{\mathcal}

\newcommand{\eat}[1]{}

\usepackage[ruled,vlined]{algorithm2e}

\usepackage{graphicx}
\usepackage{tabularx}
\usepackage{booktabs} % for much better looking tables

\usepackage{array} % for better arrays (eg matrices) in maths
\usepackage{paralist} % very flexible & customisable lists (eg. enumerate/itemize, etc.)
\usepackage{verbatim} % adds environment for commenting out blocks of text & for better verbatim
\usepackage{subfig} % make it possible to include more than one captioned figure/table in a single float

\usepackage{float}
\usepackage{placeins}
\restylefloat{figure}
\usepackage{dsfont}

\usepackage{listings}
\usepackage{amsfonts}
\usepackage{amssymb}
\usepackage[mathscr]{euscript}
\usepackage{latexsym,amsmath}
\usepackage{mathtools}
\usepackage{afterpage}
\usepackage{enumitem}

\usepackage[amsthm, amsmath, thmmarks, thref, hyperref]{ntheorem}

\usepackage{scalerel}

\usepackage{imakeidx}

\usepackage[usenames,dvipsnames,svgnames,table,xcdraw]{xcolor}
\usepackage[hyperindex, breaklinks, colorlinks, urlcolor=magenta, citecolor=red, linkcolor=blue]{hyperref}

\usepackage{tikz}
\usetikzlibrary{trees}

\usepackage{forest}
\usetikzlibrary{arrows.meta, shapes.geometric, calc, shadows}

\colorlet{mygreen}{green!75!black}
\colorlet{col1in}{red!30}
\colorlet{col1out}{red!40}
\colorlet{col2in}{mygreen!40}
\colorlet{col2out}{mygreen!50}
\colorlet{col3in}{blue!30}
\colorlet{col3out}{blue!40}
\colorlet{col4in}{mygreen!20}
\colorlet{col4out}{mygreen!30}
\colorlet{col5in}{blue!10}
\colorlet{col5out}{blue!20}
\colorlet{col6in}{blue!20}
\colorlet{col6out}{blue!30}
\colorlet{col7out}{orange}
\colorlet{col7in}{orange!50}
\colorlet{col8out}{orange!40}
\colorlet{col8in}{orange!20}
\colorlet{linecol}{blue!60}

\newcommand\boldblue[1]{#1}
\title{Adversarial Decisions on Complex Dynamical Systems using Game Theory
}

\author{
  Andrew C. Cullen \\
  School of Computer and Information Systems\\
  University of Melbourne\\
  Melbourne, Australia \\
  \texttt{andrew.cullen@unimelb.edu.au} \\
   \And
  Tansu Alpcan \\
  Department of Electrical and Electronic Engineering\\
  University of Melbourne\\
  Melbourne, Australia \\
  \And 
  Alexander C. Kalloniatis \\
  Defence Science and Technology Group\\
  Canberra, Australia\\
}

\begin{document}
\maketitle 
\title{Adversarial Decisions on Complex Dynamical Systems using Game Theory}

%Computational Approaches to a Game-Theoretic Model for Adversarial Decision Making}

\begin{abstract}
We apply computational Game Theory 
to a unification of physics-based models that represent decision-making across a number of agents within
both cooperative and competitive processes.
Here the competitors try to both positively influence their own returns, while negatively affecting those of their competitors. Modelling these interactions with the so-called Boyd-Kuramoto-Lanchester (BKL) complex dynamical system model yields results that can be applied to business, gaming and security contexts. This paper studies a class of decision problems on the BKL model, where a large set of coupled, switching dynamical systems are analysed using game-theoretic methods.

Due to their size, the computational cost of solving these BKL games becomes the dominant factor in the solution process. To resolve this, we introduce a novel Nash Dominant solver, which is both numerically efficient and exact. The performance of this new solution technique is compared to traditional exact solvers, which traverse the entire game tree, as well as to approximate solvers such as Myopic and Monte Carlo Tree Search (MCTS). These techniques are assessed, and used to gain insights into both nonlinear dynamical systems and strategic decision making in adversarial environments.

\end{abstract}

\keywords{Dynamical Systems; Game Theory; Decision Making; Monte Carlo Tree Search.}

\maketitle

%\IEEEdisplaynontitleabstractindextext

% \fi
%\IEEEpeerreviewmaketitle

\section{Introduction} \label{sec:intro}

In this paper we study through the lens of Game Theory
a complex dynamical system that
is a unification of physics-originating models but
applied to a competitive decision-making context.
The paper solves the model using modern computational methods and presents a parameter analysis to guide practical applications. The physics-based model seeks
to represent a tension that is inherent to adversarial decision making processes involving multiple agents, between cooperation and competition.
The model thus touches upon both cognitive and computational science. In order to study such processes, this paper presents a novel numerical treatment for game-theoretic solutions of large scale simultaneous move adversarial games conducted between rival agents.

In this model, each player is connected to others within their group through a nodal network structure representing agents (or subsystems) aligning with the players goals through a Kuramoto model \cite{Kuramoto1984}. This network model for oscillator synchronisation has been used as the basis of representation of a diverse set of natural, technological and social systems \cite{doerfler2014, kalloniatis2019controlsync, wu2020synchronization}. 
In this context the model is designed to represent
a continuous competitive Perception-Action cycle \cite{Neisser1976} between any two agents, 
known in some contexts as the Boyd Observe-Orient-Decide-Act (OODA) loop \cite{BoydOODA}. This model has been diversely applied to business, cybersecurity, and military contexts \cite{negash2008business, demazy2018game, andrade2019cognitive}. 
The representation of the OODA loop through the Kuramoto model has been shown to apply to both competing sets of decision-makers \cite{KALLONIATIS201621,HOLDER201710} and for decision makers acting in isolation \cite{kalloniatis2020HQ}. \boldblue{The Kuramoto model for a single group may be seen as a mathematical sociological model,
as seen in applications to opinion dynamics
\cite{pluchino2006opinion} for example. 
The competitive aspect of the two-network
variation of the Kuramoto-Sakaguchi system \cite{Sakaguchi1986} naturally
lends itself to a game theoretic treatment. 
This version thus provides a representation
of a competitive Command-and-Control (C2) context
in a more generic approach than previous
physics-based treatments \cite{SONG20135206,SONG2015322}.
The model captures two such `social' systems with cooperation sought within, and competition across, each. }

Coupling these oscillator models to a Lanchester model \cite{Lanchester1916, MorseKimball-1951,mackay2006lanchester} allows for outcomes of the decision making process to be quantified, such that success in coordinated
decision-making results in enhanced resources of one
and depletion of the resources of the competitor.
The Lanchester model is itself an adaptation of the predator-prey dynamics, or multi-species 
Lotka-Volterra model. 
\boldblue{ As a representation
of growth and decay of
entities, these models describe physical processes often obeying conservation laws,
with wide application in ecology \cite{BRADSHAW1998107}. }
The unification of these models,
Kuramoto and Lanchester, was first
proposed in \cite{Ahern2020}, and in this context is called
the Boyd-Kuramoto-Lanchester (BKL) dynamical system. 
In essence, because both models admit treatment
in continuous-time differential equations their
unification is entirely natural.

\begin{figure}[htp]
	\centering
	\includegraphics[width=0.9\columnwidth]{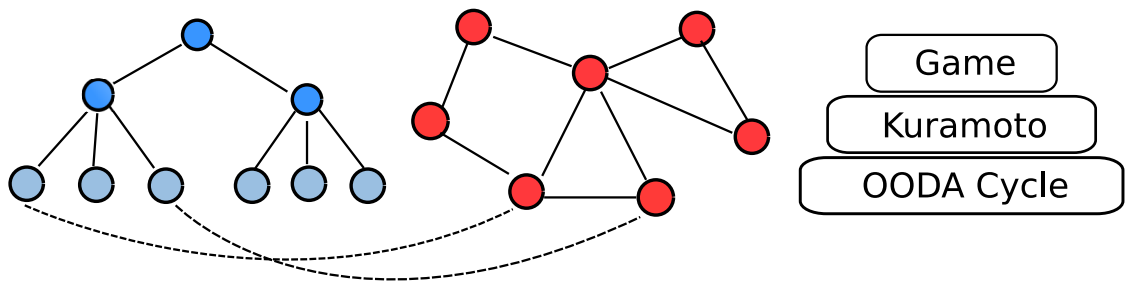}
	\caption{\boldblue{Conceptual diagram of the Boyd-Kuramoto-Lanchester (BKL) games.} Nodes represent agents, with the player corresponding to the aggregate set of nodes. Solid and dashed links respectively represent networked connections between agents and adversaries \boldblue{, and the different shades of blue indicate relative  state of synchronisation of agents} in the hierarchical structure.\label{fig:bkldiagram} }
\end{figure}

Of particular interest is the influence of different network structures on the cohesion and adaptability of players. For this work we deliberately consider an asymmetric arrangement in which one group of players is subject to a hierarchical network structure. The other group has a more organic and interlinked network topology coupling its oscillators. These structures are  represented respectively by the Blue and Red players of Figure~\ref{fig:bkldiagram}. The interactions between these two groups of players is fixed such that only a subset of the nodes from each directly interacts with those of the opponent group; nodes not connected in this way may be said to play leadership roles within the group. 

This model of two players engaged in adversarial decision making under constrained resources rewards internal synchronisation, but also incorporates the potential for adversary driven outcomes that undermine the capacity for coupling. The competitive nature of these components creates a process that is inherently Game-Theoretic in nature \cite{Ahern2020}, and exists alongside other recent works tying dynamical systems to Game-Theory \cite{li2020exploring}. Due to their successful application to multiple adversarial environments, the mathematical and conceptual framework of Security Games \cite{alpcan-book} is applied to two-player adversarial BKL games. When the outcome of these systems are determined by a multi-stage decision making process, these games present an as yet unexplored challenge, in terms of both their analysis and the development of appropriate solution strategies under computational constraints. As such, particular focus is placed upon both establishing the theoretical basis for such a multidisciplinary framework, and developing and implementing computational tools that are suitable for such a model.

The \textbf{contributions} of this paper include:
\begin{itemize}
 \item Constructing a novel union of dynamical systems and game theory through the BKL model of networked oscillators.
 \item Introducing novel numerical algorithms for solving game theoretic problems, with a focus upon numerical scaling and large game trees. 
 \item Detailed numerical analysis of the outcomes of game theoretic dynamical systems, with a particular focus upon understanding asymmetric adversarial decision making processes, which provides insights for practical applications.
 \eat{
 \item Developing an understanding of asymmetric adversarial decision making games. 
 \item Developing novel algorithms for obtaining Nash equilibrium solutions of discrete dynamic game formulations of the model in extensive form, with a focus upon large game trees.
 \item A detailed numerical analysis of the game outcomes investigating the impact of system and network parameters on Nash Equilibrium solutions under different scenarios.
 }
\end{itemize}
To support this, the paper begins by introducing the dynamical systems model for BKL dynamics. Following this,  Section~\ref{sec:gamemodel} introduces a specific game-theoretic formulation. To facilitate the solution of such games, a range of computational techniques to solve the discrete dynamic games is presented in Section~\ref{sec:solutiontechniques}.
The behaviour of the game solutions to various
parameters under different scenarios is discussed in Section~\ref{sec:sensitive}. The paper concludes with remarks and a discussion on future
research directions.

\section{Boyd-Kuramoto-Lanchester Complex Dynamical Systems} \label{sec:boydmodel}

In the following we present first the deterministic two-network Kuramoto-Sakaguchi \cite{Sakaguchi1986} oscillator model, and discuss how it is mapped to an adversarial context; at this level the representation is called the `Boyd-Kuramoto' (BK) Model as it captures competing OODA loop cycles as a continuous process in the phase oscillator at the heart of the formulation. Next, we incorporate into this the well-known Lanchester model to provide the combined BKL Model. This summarises the original
proposal in \cite{Ahern2020}.

\subsection{Boyd-Kuramoto Dynamical Model} \label{sec:bkmodel}

Let $\mc B=\{1,\ldots,N\}$ and
$\mc R=\{1,\ldots,M\}$ be the respective sets of Blue and Red Agents. Each Blue Agent $i \in \mc B$ 
has a frequency $\omega_i$ and phase $\beta_i$, and similarly each Red Agent $j \in \mc R$ has
frequency $\nu_j$ and phase $\rho_j$. The Blue Agents are connected to each other through a symmetric $N \times N$ adjacency matrix $B$; while the Red Agents via the  $M \times M$ matrix $R$. 
The  $N \times M$ matrix $A$ represents the unidirectional external links from Blue to Red Agents. While asymmetric interactions are available, for this work we impose that the interactions from Red to Blue are simply the transpose of $A$. Figure~\ref{fig:bkldiagram} visualises one possible configuration,
\boldblue{where common shades of blue (for the hierarchical
group of players) indicate agents
close in synchronisation to each other}. The quantities $\zeta_B$, $\zeta_R$, $\zeta_{BR}$, and $\zeta_{RB}$, are respective coupling constants for Blue and Red internally and for Blue to Red and {\it vice versa}.
The resulting Boyd-Kuramoto model is inherently nonlinear, and admits complex and chaotic dynamics that can be derived through (typically numerical) solutions of
\begin{align} \label{e:kuramoto1}
 \frac{d \beta_i}{dt}  = & \; \omega_i - \zeta_B \frac{\sum_{j \in \mc B} B_{ij} \sin(\beta_i - \beta_j)}{\sum_{j \in \mc B} B_{ij}} \nonumber\\ 
 & - 
 \zeta_{BR}\sum_{j \in \mc R} A_{ij}  \sin(\beta_i - \rho_j - \phi), \;\; i \in \mc B \\
 \frac{d \rho_i}{dt}  = & \; \nu_i - \zeta_R \frac{\sum_{j \in \mc R} R_{ij} \sin(\rho_i - \rho_j)}{\sum_{j \in \mc R} R_{ij}} \nonumber \\
  & -
 \zeta_{RB}\sum_{j \in \mc B} A^T_{ij}  \sin(\rho_i - \beta_j - \psi), \;\; i \in \mc R, \nonumber
\end{align}
where $(\cdot)^T$ is the transpose operator, and $\phi$ and
$\psi$ are the phase lags (frustrations) \cite{KALLONIATIS201621}. These two lags capture the essence of Boyd's
proposal that advantage is sought by one side over the other insofar as the coupled dynamics influence the realisation
of one side being {\it ahead} of the other by the desired amount: $\phi$ for Blue, and $\psi$ for Red. Whether collectively the intended `aheadness' of one or the other is achieved depends on the evolution of the non-linear dynamics.

\subsection{Boyd-Kuramoto-Lanchester Dynamical Model} \label{sec:bklmodel}

To extend the set of admitted dynamics, a Lanchester model of adversarial interactions can be incorporated within the model, in order to quantify the implications of the decision making processes. The combined Boyd-Kuramoto-Lanchester (BKL) model is immediately applicable to competitive decisions as it captures the complex cyclic decision processes and their effects on adversarial populations of networked heterogeneous agents, in a manner that builds complexity through the aggregated model dynamics. 

%In many contexts, the outcomes of the decision making framework, as represented by the Boyd--Kuramoto model, exist as part of a broader adversarial formulation.

To this point the Boyd-Kuramoto decision making is detached from any outcomes \boldblue{as might be
realised in the physical state of the entities}. Representing the outcomes of the competitive process can be achieved by coupling the Boyd-Kuramoto equations to a larger dynamical system. Some options for these include Colonel Blotto Games \cite{Roberson06}, Volley/Salvo models
\cite{Hughes95}, and the Lanchester model  \cite{Lanchester1916}. Of these, the Lanchester model holds particular interest, due to well understood competitive properties and applicability to Operations Research. For this the resources---or force strengths---of players $R$ and $B$ are quantified by
\begin{equation}\label{eq:BKL}
\frac{dP_{R}}{dt} = -\alpha_{BR} P_{B} \hspace{1 cm} \text{and} \hspace{1 cm} \frac{dP_{B}}{dt} = -\alpha_{RB} P_{R},
\end{equation}
%\begin{align}\label{eq:BKL}
%\frac{dP_{R}}{dt} &= -\alpha_{BR} P_{B},\\
%\frac{dP_{B}}{dt} &= -\alpha_{RB} P_{R}, \nonumber
%\end{align}
where $\alpha_{BR}$ and $\alpha_{RB}$ are relative measures of adversarial effectiveness between the respective agent populations. When $(\alpha_{RB}, \alpha_{BR})$ are constant the Lanchester equations are integrable and admit a unique solution. 

In an adversarial environment it is reasonable to expect that each $\alpha$ is no longer strictly constant, but rather exhibits a dependence upon the effectiveness of
the made decisions. As such, the full BKL model takes the form
\begin{align} \label{e:bklfull}
\frac{dP_B}{dt} = &  - \kappa_{RB} \cdot \frac{\abs{\sum e^{j\rho_i}}}{N_r} \cdot \frac{\sin(\frac{\sum \rho_i}{N_r}-\frac{\sum \beta_i}{N_b})+1}{2} \cdot P_R, \\
\frac{dP_R}{dt} = & - \kappa_{BR} \cdot \frac{\abs{\sum e^{j\beta_i}}}{N_b} \cdot \frac{\sin(\frac{\sum \beta_i}{N_b}-\frac{\sum \rho_i}{N_r})+1}{2} \cdot P_B,  \nonumber
\end{align}
with populations thresholded to prevent physically infeasible negative populations. This equation is coupled to  (\ref{e:kuramoto1}) through $(\beta_i, \rho_j)$, and $N_r=M$ and $N_b=N$ correspond to the cardinalities of the Blue and Red agent sets. %, and the phases $\beta_i$ and $\rho_j$ respectively represent the $i^{th}$ and  $j^{th}$ Blue and Red Agents are determined by (\ref{e:kuramoto1}). 
Note that this model may be called a global model,
where the resources of the two sides are homogeneous.
In \cite{Ahern2020} a heterogeneous form of the model is
also given, where the resources may also be structured
through network parameters
using the generalisation of the Lanchester model
in \cite{Kalloniatis2020NetLanch}. We do not treat this model here in this first application of computational game Theory to such a system.

\section{A Dynamic Game Approach to Competitive Decisions on Complex Systems}\label{sec:gamemodel}
In the above model, one may derive thresholds or optima for decisions by one side assuming fixed parameters for the other, as in \cite{KALLONIATIS201621}.
The decisions here are about choices of one or more of network structure, couplings, frequency lay-down or degree of aheadness.
The fact
that the competitor has a say in the outcome
(success of those choices) means that a game-theoretic treatment is essential. Our work follows a previously developed framework 
\cite{basargame}, in which the BKL engagement can be classified as a two-player, non-zero sum, strategic game. Within this context, the players control their own sets of networked or connected populations of agents $\mc R, \mc B$, for $Red$ and $Blue$ respectively, along with their corresponding adjacency matrices $R, B$, representing the underlying connection graphs, a conceptual representation of which is shown in Figure~\ref{fig:bkldiagram}. These graphs play a significant role within both BK and BKL models as captured by the dynamic equations (\ref{e:kuramoto1}) and (\ref{e:bklfull}), respectively.

The actions of the players can be considered as control or input variables of the BK and BKL models. As a specific choice, we assume that the players decide on their strategic goals of leading or lagging targets $\phi \in [0, \pi]$ and $\psi \in [0, \pi]$ respectively, representing their desired position in the Boyd (OODA) cycle, as described in the BK and BKL equations (\ref{e:kuramoto1}) and (\ref{e:bklfull}). %After each decision point, 
At discrete decision points in time $T_i$,
which subdivide the overall engagement time $[0,T_f]$,
the BKL equations are numerically integrated over a finite time horizon $t \in [T_{i},T_{i} + \delta t]$, where $\delta t$ is long enough to allow for the system dynamics to meaningfully evolve over the decision window. Over this time horizon, the players
decide on their actions in terms of the strategy vectors
\begin{equation} \label{e:playerstrategies}
S_{B} = [\phi_0, \phi_1, \ldots,   \phi_K], \hspace{0.75 cm} \text{and} \hspace{0.75 cm} 
S_{R} = [\psi_0, \psi_1, \ldots,   \psi_K].
\end{equation}
Over the finite set of time steps $t_i$,  $i=1,2,\ldots, K$, corresponding to stages in the time interval $t \in (0, T_f )$. The resulting game is formally defined by the tuple $\mc G_{BKL,dyn}:=\langle \mc P, (S_B,S_R), (U^d_B,U^d_R) \rangle,$ \boldblue{where $\{U^d_B, U^d_R\}$ are the utilities of the Blue and Red players after $d$ actions have occurred, and $\mc P=\{Blue, Red\}$ represents the set of players.} 

Each round of the game corresponds to a level of the game tree (extensive form) starting from the root node on top. Under the assumption that the choice of $(\phi, \psi)$ is sampled from a discrete action-space---which allows the game to be considered in a computationally tractable manner---each level of the game then is a static bi-matrix game with utility values and players actions dictated by the underlying dynamic BKL model. This is not a repeated game since the underlying game state changes with each round or as we go down the tree level by level as a result of BKL dynamics. 

As the game evolves, the game tree reaches a terminal state when either a fixed time point $T_f$ has been reached; or that one player breaches a termination condition. This termination condition can be flexibly defined, but in the BKL context typically would be that a player depleted in resources to the degree that they no longer have the capability to effectively participate within the game. At the games end state the outcome is quantified by a pair of utility functions measuring the final balance of resources
\begin{equation} \label{e:staticbklutils}
U^s_{B}(T_f ) = P_{B}(T_f ) - P_{R}(T_f ), \hspace{0.75 cm} \text{and} \hspace{0.75 cm}
U^s_{R}(T_f ) = P_{R}(T_f ) - P_{B}(T_f ).
\end{equation}

The game-theoretic model of the players' dynamics assumes that the agents are rational decision makers who are choosing strategies and taking actions that will maximise their own utilities, in light of the predicted and observed behaviour of their opponent. Such decisions also must consider the order of play, which can be either simultaneous or sequential, with the latter leading to a `Stackelberg` (or `Leader-Follower') game structure. We focus on each player taking actions concurrently, in which neither player has an information advantage over the other at each decision point. Such a game structure is particularly well suited to high tempo decision making environments similar to the original military context of the OODA loop. It is worth noting that players having this information does not necessarily mean that they have computing power to calculate the entire game tree, in other words all possible outcomes of the game. This combinatorial complexity distinguishes the game at hand from classical full information games~\cite{basargame}. 

We use the deterministic pure strategy classical Nash Equilibrium (NE) as the solution concept to explore the behaviour of the agents and their optimal behaviour within the competitive environment. Formally, the NE is the set of player strategies (and associated utilities) where no player gains deviating from their strategy, when all other players also follow their own NE. It can corresponds to fixed point and the intersection point of players best responses \cite{basargame}. It is worth noting that bi-matrix games that are solved at each level always have a solution in mixed strategies, corresponding to a probability distribution over the actions (pure strategies) \cite{basargame}. The choice of pure strategies---in contrast to probabilistic mixed strategies---reflects the low likelihood of repeated replay for BKL scenarios of interest.

However, pure-strategy NE may not exist in the class of games considered here, and as such it is natural to consider the security strategies of players, which ensure a minimum performance. Also known as minmax and maxmin strategies, these strategies allow each player to establish a worst-case bound on minimum outcome~\cite{gtessentialsbook}. However, in the absence of a NE solution, these can be overly conservative, which has led to alternative solution concepts such as regret minimisation. Another related solution concept is the $\epsilon-$NE, which is an approximation to NE solution~\cite{gtessentialsbook}.

\section{Strategic Solution Techniques} \label{sec:solutiontechniques}

Solving a dynamic game formulated as a BKL complex system, and hence obtaining best response strategies of players, involves both constructing Game Theoretic solutions to the overall game, and the numerical solutions of the BKL that the Game Theoretic results depend upon. As the BKL ordinary differential equations (ODEs) are constant coefficient, coupled initial value problems with trigonometric nonlinearities, solving these differential equations is a relatively straightforward process and is standard in studies of the Kuramoto model for complex networks, however, inherently this process becomes a hurdle as the size of the game tree increases. 

In order to consider large-scale decision processes while being cognisant of computational constraints, we consider approximate solutions of the BKL equations using the Dormand-Price Runge-Kutta method \cite{dormand1996}, using a coarse fixed step-size. While the coarse step-size results in inaccurate results in terms of player utilities, our investigation has shown consistency between the optimal player strategies in the cases of more accurate and approximate solvers, even when the player utilities deviate from each other. Making this change significantly decreases the overall computational cost, without influencing the relative cost across solution methodologies, allowing the scaling properties of each algorithm to be assessed. 

\begin{center}
    \scalebox{0.7}{
    \begin{minipage}{0.9\linewidth}
\begin{algorithm}[H]
        \While{Exploring Tree}{
        \While{Depth $<$ Terminal}{
		Identify available actions\;
		Select action and save action to path\;
        }
        Solve game for path\;
        Backpropagate information\;
        }
        \caption{Generalised Process for Tree Exploration}
        \label{alg:general}
        \end{algorithm}
\end{minipage}
}
\end{center}

\subsection{Full Competitive Decision (Game) Tree Solver}

The Full Tree solver operates by constructing an extensive form representation of the competitive decision process, comprising all potential choices of the action parameters $\phi$ and $\psi$ at each decision point. Performing a depth first search across all terminal leaf states, and then backpropagating the NE solution at each depth recursively from the terminal depth to the root node, yields a NE utility which represents the utility at the game's terminal state~\cite{gtessentialsbook}, that corresponds to an \textit{exact} solution to the game tree. This process directly follows Algorithm~\ref{alg:general}, where backpropagation only occurs when all potential action pairs $(\phi, \psi)$ from a point in the tree have been explored. When this condition has been met, and solved for, at the root node of the game tree, the game has been solved. 

A saddle point, at which the exact NE would sit, is not guaranteed to exist for two-player simultaneous-move zero-sum games. Therefore the NE is approximated by adopting the security (or max-min and min-max) strategies of the players~\cite{gtessentialsbook} as the solution concept by 
\begin{equation}\label{eqn:NE}
\max_{a \in \mathcal{A}} \min_{b \in \mathcal{B}} U(a,b),
\end{equation}
where $U(\cdot, \cdot)$ is a matrix of the recorded utilities corresponding to each unique decision pairing. The components $a \in \mathcal{A}$ and $b \in \mathcal{B}$ correspond to the decisions and decision sets for each player at the currently explored component of the game--tree.

While any game tree corresponding to a zero-sum simultaneous-move game can be exactly solved in this manner, the computational burden of resolving all possible game states can make this process intractable. This limitation stems from the polynomial growth of the size of the game tree as the number of action states and decision points are increased. If $|\mathcal{A}|$ and $|\mathcal{B}|$ represent the size of the action space for each player, and $d$ the number of decision points, then the size of the game, and the ensuing computational cost can be shown to be $\mathcal{O}\left(|\mathcal{A}|^{d} |\mathcal{B}|^d\right)$. For future reference, we shall impose that $|\mathcal{A}| = |\mathcal{B}| \equiv B$, allowing the cost to be expressed as being $\mathcal{O}(B^{2d})$.

In the context of BKL games, the problematic nature of this growth is not a direct consequence of the size of the game trees themselves, but rather how many times the %system of BKL 
ODEs need to be solved, as it is this part of the process that dominates the computational cost. As a consequence of this, the growth of computational complexity with $\mathcal{O}(B^{2d})$ when constructing a solution using the Full Tree approach is infeasible when the game involves large decision spaces, or long time horizons involving multiple decision points.

\subsection{Nash Dominant Game Pruning}

In sequential move games, where player decisions follow one another in a sequential way, large decision trees can be solved by employing Alpha-Beta pruning, which is frequently employed to reduce the portion of the game tree that needs to be explored to reach a solution. In the best case, Alpha-Beta pruning can reduce the computational cost from $\mathcal{O}\left(B^{d}\right)$ to $\mathcal{O}\left(\sqrt{B^{d}}\right)$, by excising subgame branches that can be proven to not contain the NE, without needing to explore the subgame in question. Importantly, Alpha-Beta pruning provably preserves the game tree, thus it is classified as an \textit{exact} solver, resulting in the same NE from solving over the full tree. 

While Alpha-Beta pruning is a powerful approach for sequential move games, it can not be applied to simultaneous move games. As such, we present \textit{Nash Dominant Game Pruning}, a novel solution concept. Similar to Full Tree, the game is explored through a recursive process with utilities calculated at the leaves and then back-propagated up the tree. In contrast to Full Tree, Nash Dominant identifies action pairs $(\phi, \psi)$ that are strategically dominated---that will not correspond to the equilibrium state---and truncates all subsequent states within the game tree. 

This process is performed by describing the action pairs as a matrix game and identifying if any recorded utility within an incompletely explored column is smaller than the largest column minima of any completely explored column. If the observed value is smaller than the largest column minima then it follows by Equation (\ref{eqn:NE}) that the NE cannot exist inside that column. As such, the subgames that stem from those points in decision space can be excluded.  The totality of this process only affects the process of selecting available actions within Algorithm~\ref{alg:general}, however for completeness Algorithm~\ref{alg:NDsolver} presents all steps for Nash Dominant Game Pruning. For this, evaluating the path refers to finding the terminal state utility from the state of the two players, and scoring the result involves solving Equation (\ref{eqn:NE}). 

\begin{center}
    \scalebox{0.7}{
    \begin{minipage}{0.9\linewidth}
\begin{algorithm}[H]
\KwData{NashDominant(P, $i$, $j$, $min_c$ , D) for P being the historic path, $(i, j)$ representing the decision index from $(\phi, \psi)$, and D is the Depth into the game tree.}
Path = Path + (i, j)\;
\eIf{Depth $<$ Terminal Depth}{
	\While{Any element of the decision space is unexplored}{	
			\For{All (i,j) action pairs in col that have not been explored}{
				results(i,j) = NashDominant(P, i, j, $min_c$ , D + 1)\;
				\If{results(i, j) $<$ $min_c$}{
					Set all remaining elements in the column to results(i,j)
				}
				}
				\If{min of col $>$ $min_c$}{$min_c$ = min of column}
Increment row, increment column;
}
\Return Score(results)
}
{
	\Return Evaluate(Path)
	}
\caption{Psuedocode algorithm for the Nash Dominant solver.} \label{alg:NDsolver}
\end{algorithm}
\end{minipage}
}
\end{center}

This process provably yields the NE solution if it exists, and the security strategy if it does not. Under the best case scenario, the algorithm would only need to solve $2 B - 1$ subgame states from each game state, with $\mathcal{O}(B^{d})$ complexity. At worst case this scheme scales with $\mathcal{O}(B^{2d})$.

\subsection{Myopic Approximation}

One advantage of games like the BKL model is that the utility corresponding to a state can be exactly calculated, even if it does not correspond to a terminal node. This is a consequence of the game being scored in terms of the changes in Lanchester resources or force strengths, rather than a binary win-loss state.  As such, an approximate solution for large game trees can be found by evaluating the utility resulting from all action pairings at the first depth, and then selecting the equilibrium solution---effectively repeated calls of Algorithm~\ref{alg:general} for a game depth of $1$. At all subsequent depths, the tree is pruned so only actions stemming from the equilibrium solution at the previous depth are considered. Repeating this process to the terminal depth of the tree results in what is known as the \textit{Myopic} (or \textit{Greedy}) \textit{approximate} solution. The validity of this approach is built upon the premise that the influence of early decisions likely dominates the NE, due to the dependence of system outputs on the Lanchester exponential decay. Under this premise, Myopic search can yield an accurate and numerically efficient estimate of the NE through a limited breadth-first search scaling with $\mathcal{O}(d B^{2})$. However, the nonlinear nature of the BKL system indicates the potential for the system to exhibit chaotic dynamics, contradicting the assumptions of Myopic search.

\subsection{Monte Carlo Tree Search (MCTS) \label{Sec:MCTS}}

Monte Carlo techniques are well known for their ability to numerically approximate solutions through partial exploration of large and complex spaces. In the case of tree structures, Monte Carlo Tree Search (MCTS) encompasses a family of techniques for estimating the NE of sequential or simultaneous move games. The utility of MCTS for complex games has been demonstrated in a number of board games \cite{gelly2012}, as well as more general scenarios for decision making in complex environments \cite{arneson2010, browne2012, lanctot2013, chen2015decentralized, haeri2017virtual}.

Constructing a solution to a game with MCTS requires striking a balance between exploitation of the game tree, in which visiting new subgame regions are prioritised, and exploitation of previously visited subgame regions,  in order to improve knowledge about their subgame states. MCTS provides a framework for iteratively exploring game trees, with the aim of reaching an equivalence class with the full game tree. 
As the MCTS procedure explores the game-tree, the observed game structure asymptotically converges to that of the full game. Under the restriction that the selection approach is $\epsilon$-Hannan consistent, MCTS algorithms converge to an approximate NE of the extensive-form game \cite{lisy2013}. Our investigation is based upon the Decoupled UCT (DUCT) algorithm, which balances exploration and exploitation of the game tree in a fashion inspired by multi-armed bandits  \cite{kocsis2006}. After each player has visited each possible action once, subsequent actions chosen by
\begin{equation}\label{eqn:MCTSargmax}
a_{i} = \text{argmax}_{a \in \mathcal{A}_{i}(s)} \left\{ \frac{f(X^{i}_{s,a})}{n_{s,a}}+ C \sqrt{\frac{\ln{n_s}}{n_{s,a}}} \right\}.
\end{equation}
Here $X_{s,a}$ is the sum of rewards for the subgame corresponding to the action $a$ from state $s$, corresponding to the position in the partially explored game tree, and both $n_{s}$ and $n_{s,a}$ represent the number of visits of the current game state, and of each of its subgame actions respectively. The function $f$ is a departure from previous implementation of DUCT, and serves to map $X$ to within $[0,1]$, based upon the largest and smallest rewards that have been observed to that point within the MCTS exploration process, across all visited leaf nodes, in order to make MCTS more appropriate for games with a scored output, rather than binary win-loss states. Thus the first term of Equation~\ref{eqn:MCTSargmax} rewards exploiting areas of the game tree that are known to be of high utility for each player, and the second biases the search process towards actions for which less knowledge about the terminal states is available. 

\begin{center}
\scalebox{0.7}{
\begin{minipage}{0.9\linewidth}
\begin{algorithm}[H]
\KwData{MCTS(P, node, $i$, $j$, $a_1, a_2$, D) for P being the historic path, $(i, j)$ representing the decision index from $(\phi, \psi)$, and D is the Depth into the game tree.}
Path = Path + (i, j)\;
\If{node is a terminal state}{
	\Return the node evaluated at that point\;
}{
\If{node is not terminal and Expansion criteria is met}{
score = MCTS(P, node, $i$, $j$, $a_1, a_2$, D)\;
Update(node)\;
\Return score\;
}
{$a_1, a_2$ = Selectively Exploit node
Update(node)\;
	\Return score
}
}
\caption{MCTS DUCT approximate game solver psuedocode algorithm.} \label{alg:mctsduct}
\end{algorithm}
\end{minipage}
}
\end{center}

Algorithm~\ref{alg:mctsduct} utilises backpropagation, in a similar fashion to the Full Tree and Nash Dominant solvers, to pass leaf information up the tree. However, due to memory concerns with large trees, the update procedure only adds to the average reward in each of the parent game states of a leaf node, as stored in $X^{i}_{s,a}$. Due to both this and MCTS's lack of guarantees for fully exploring subgames directly solving the NE is not possible, and would require an expensive leaf-to-root approach. Instead, the NE can be approximated following a top-down path by search heuristics \cite{coulom2006, chaslot2008, schadd2009}. While many such approaches exist, our approach selects the action with the greatest expected utility by solving Equation~\ref{eqn:MCTSargmax} for $C=0$.

\section{Numerical Analysis of BKL Game Solutions}\label{sec:NumericalAnalysis}

\begin{figure}
\centering
  \includegraphics[width=0.7\columnwidth]{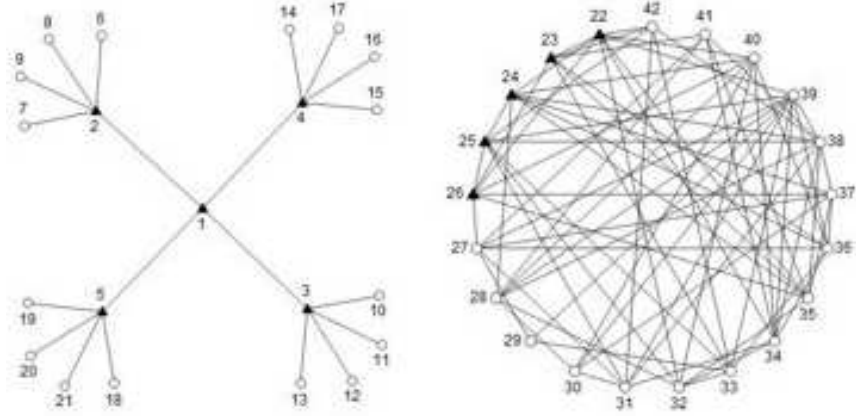}
  \caption{\boldblue{Specific network structures for the Blue (Hierarchical, left) and Red (Random, right) players, as used in all the following experiments.}}
  \label{fig:ForceStructures}
\end{figure}%

\begin{figure*}[htp]
\centering
%\begin{minipage}{.5\linewidth}
\subfloat[Population difference]{\label{equilibrium:b}\includegraphics[scale=.35]{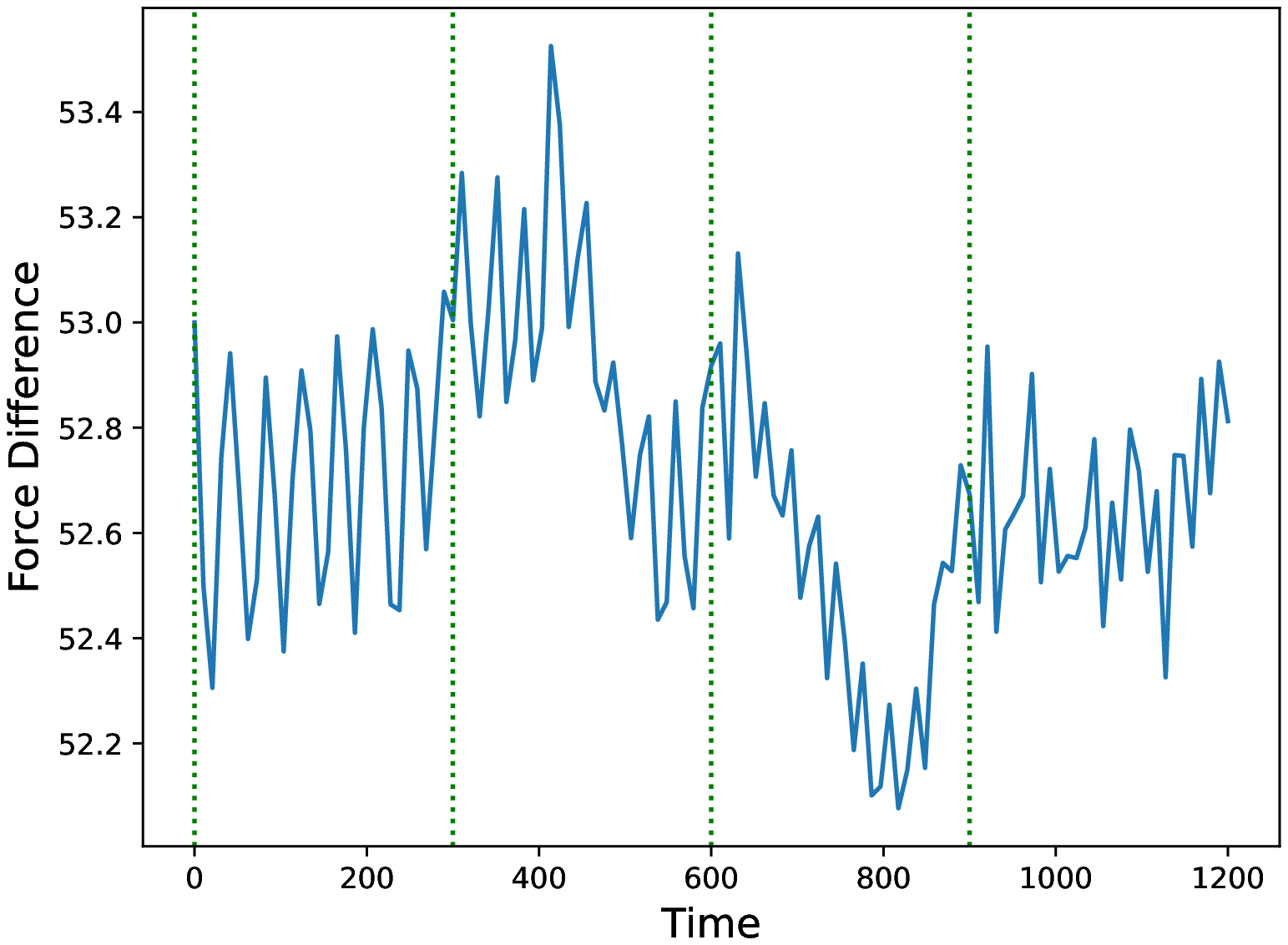}}
\subfloat[$\phi$ (Blue) and $\psi$ (Red), radius increasing with time]{\label{equilibrium:c}\includegraphics[scale=.4]{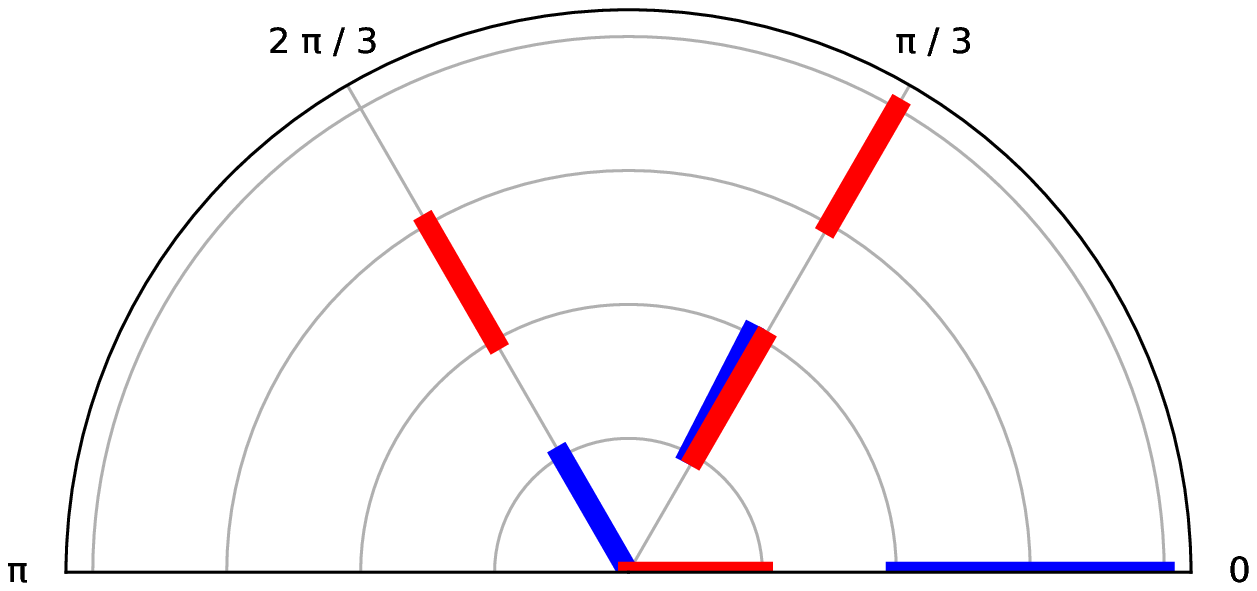}}
\caption{Force strength difference (a) and actions $(\phi, \psi)$ in (b) \boldblue{for the Red and Blue players (as per Figure~\ref{fig:bkldiagram})} when following the NE. Game has four potential actions at each of four decision points\boldblue{, with each of the decision points being denoted by the vertical green lines in panel (a)}. Parameters according to Table~\ref{Table:Parameters_1}}
\label{fig:equilibrium_solution_47}
\end{figure*}

For a game where each player has the choice of four actions across $[0, \pi]$, Figure~\ref{fig:equilibrium_solution_47} outlines the solved equilibrium dynamics. In this example, each player is able to re-position at $t = \{0, 300, 600, 900\}$, and follows the parameters of Table~\ref{Table:Parameters_1} and force structures of Figure~\ref{fig:ForceStructures}. The asymmetry between the players is deliberate, and has been chosen to produce a closely balanced competitive environment between the two players with distinct organisational structures. As will be discussed later in this chapter, the interplay between the structure of the force and its strength relative to their opponent can be assessed by perturbing the parameters of this balanced environment. 

\begin{table}[h]
\centering
\scalebox{0.7}{
\begin{tabular}{|l|c|}
\hline
\textbf{Parameter} & \textbf{Value} \\
\hline
Initial population (B and R) & $100$ \& $47$ \\ 
mean($\omega_{i}$) & $0.5032$ \\ 
mean($\nu_{j}$) & $0.5513$ \\
$(\zeta_{B}, \zeta_{R})$ & $(0.5, 0.5)$ \\ 
$(\zeta_{BR}, \zeta_{RB})$  & $(0.4, 0.4)$ \\ 
$(\kappa_{BR}, \kappa_{RB})$ & $(0.005, 0.005)$ \\
$(\epsilon_{1}, \epsilon_{2})$ & $(10^{-15}, 10^{-20})$ \\
$(\gamma_{B}, \gamma_{R})$ & $(10^{-3}, 10^{-5})$\\
\hline
\end{tabular}
}
\caption{BKL parameter space used in Section~\ref{sec:NumericalAnalysis}, reflecting recent practical examples \cite{KALLONIATIS201621,HOLDER201710}. Blue and Red players respectively representing a hierarchical and peer-to-peer randomised decision making structures.}
\label{Table:Parameters_1}
\end{table}

Simulations following these parameters demonstrate that the initial force strength advantage is preserved across time (Figure~\ref{equilibrium:b}), reflecting that the population asymmetry balances out the structural differences between the two players internal organisation, allowing Red to remain competitive with Blue. However the difference between force strengths across each decision point is not conserved: there are distinct deviations at the decision points, seen in panel (b) so that at each such point Blue and Red are positioning themselves to optimise their final utility. These choices reflect the changes in each players positioning in $(\phi, \psi)$, as observed within Figure~\ref{equilibrium:c}, which shows Reds need to re-position their behaviour more frequently. From this we infer that at each decision point after the opening state Red seeks to position itself ahead in phase of Blue. However, through its initial force advantage Blue is able maintain its relative advantage to Red without this phase advantage. 

\begin{figure*}
\centering
%\begin{minipage}{.5\linewidth}
%\centering
\subfloat[Force strengths]{\label{non_equilibrium_solution:a}\includegraphics[scale=.35]{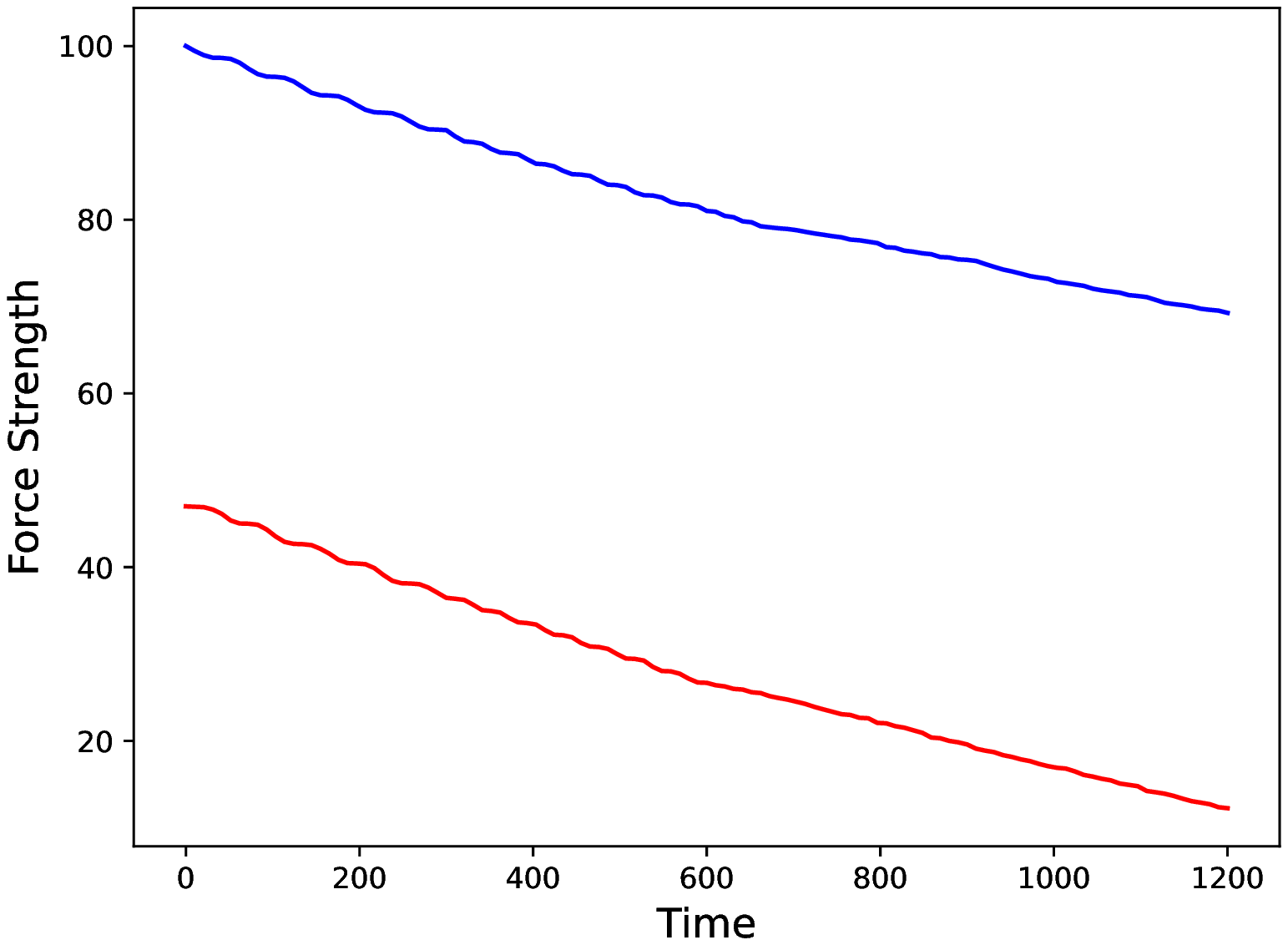}}
%\end{minipage}% \par\medskip
%\begin{minipage}{.5\linewidth}
%\centering
\subfloat[Population Difference]{\label{non_equilibrium_solution:b}\includegraphics[scale=.35]{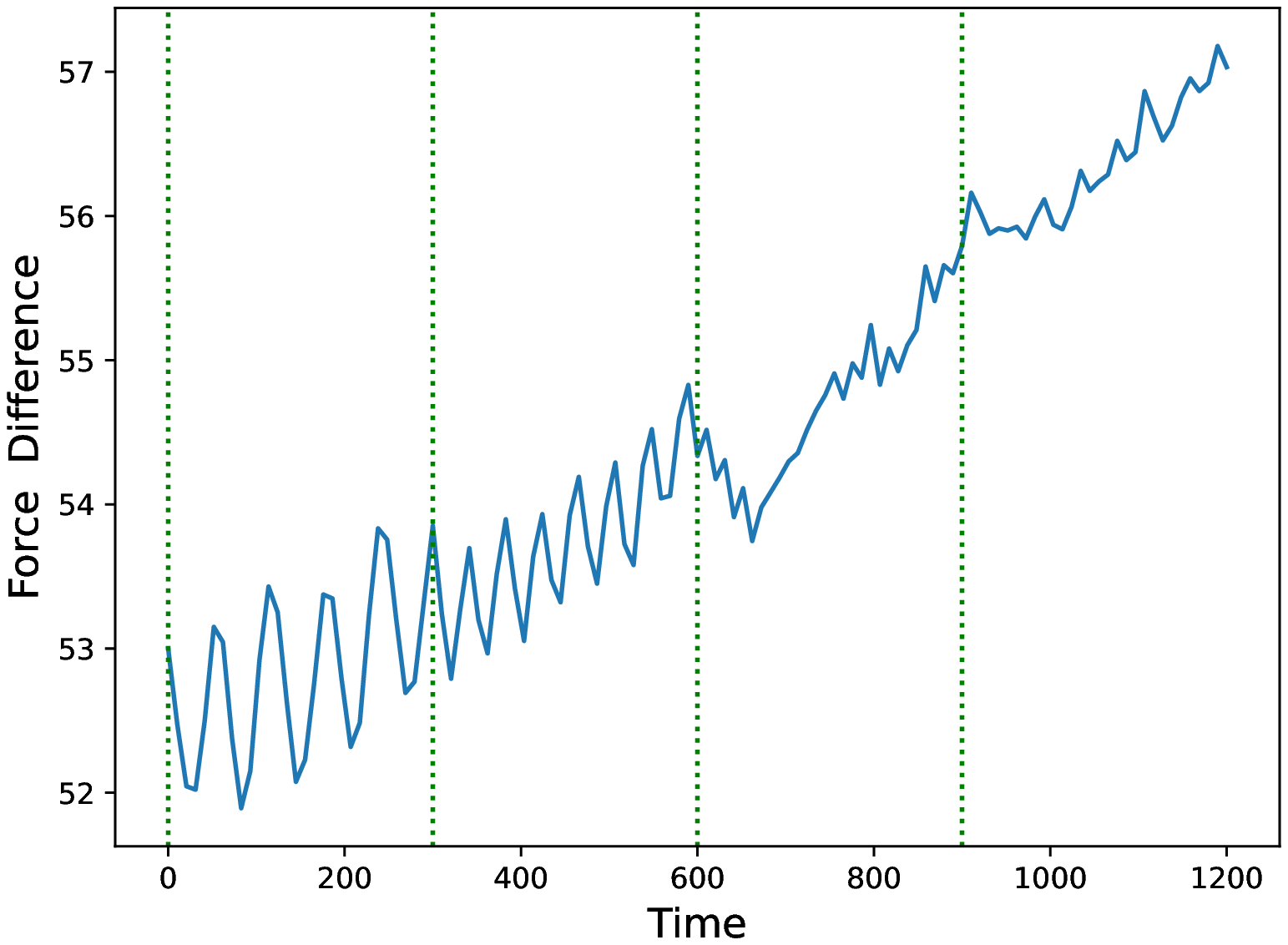}}
\caption{\boldblue{Revised player outcomes for the Red and Blue players (total force for each side in (a) and difference in (b)) when only the Blue player follows the NE strategy, based upon} Figure~\ref{fig:equilibrium_solution_47}.}
\label{fig:non_equilibrium_solution}
\end{figure*}

The nature of these outcomes are intrinssically linked to the game-theoretic solution of the game. Figure~\ref{fig:non_equilibrium_solution} demonstrates the evolution of the game if the Blue player maintains their Nash Equilibrium strategy, while the Red player deviates from the equilibrium strategy by selecting $\psi = \pi$ at each of the $4$ decision points. This change induces a deleterious outcome for Red relative to the equilibrium utility. Under the precepts of a Nash Equilibrium, the behaviour of the Blue player is not the optimal response to Red's sub-optimal (``irrational") decisions, but rather that Blue cannot be disadvantaged. The difference between the dynamics is persistently greater than $\pi/2$, which indicates that no phase locking occurs \cite{KALLONIATIS201621}.

\begin{figure*}[htp]
\centering
\subfloat[Population Difference]{\label{equilibrium100:b}\includegraphics[scale=.35]{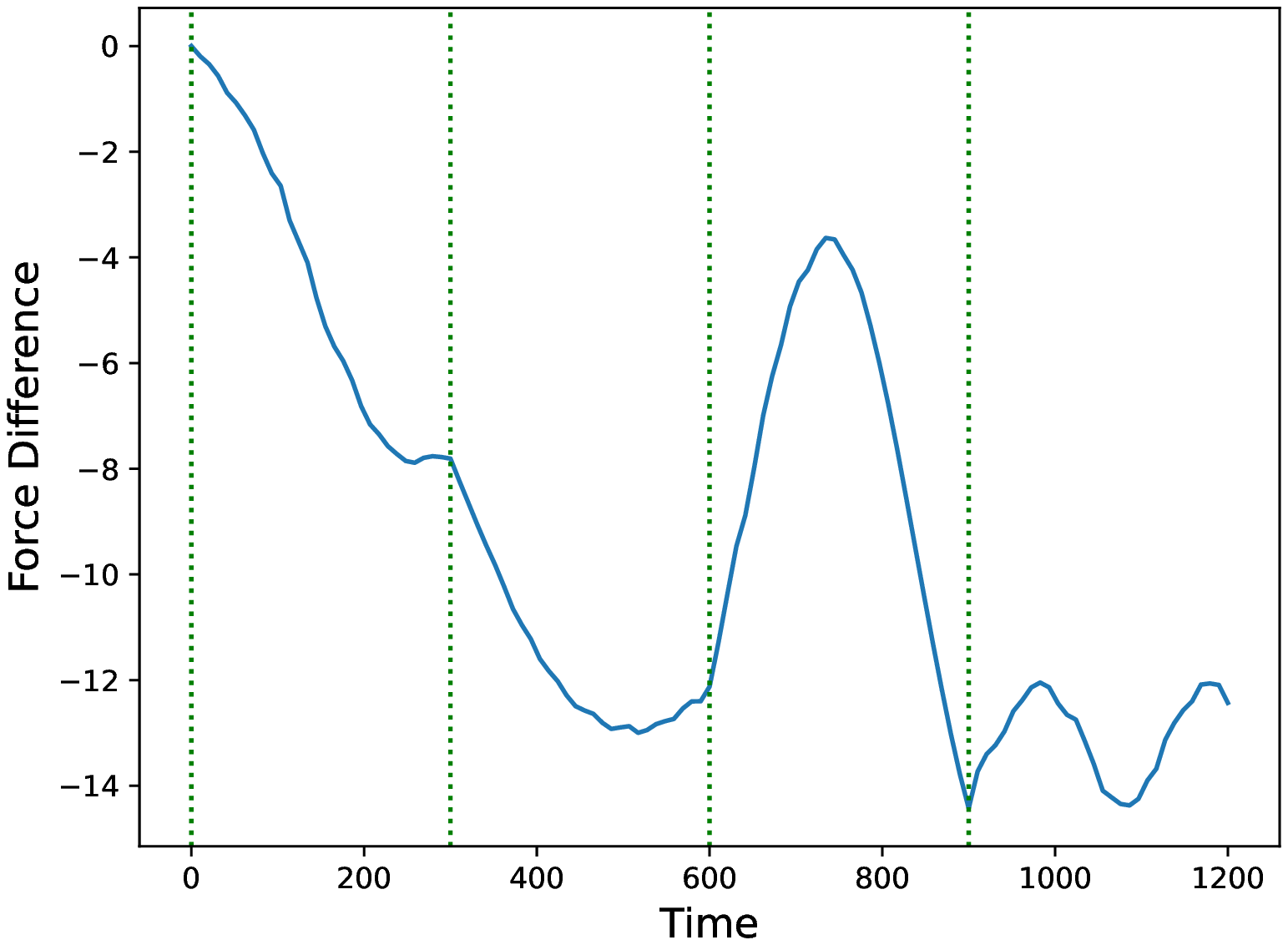}}
%\end{minipage} %\par\medskip
%\centering
\subfloat[$\phi$ (Blue) and $\psi$ (Red), radius increasing with time]{\label{equilibrium100:c}\includegraphics[scale=.4]{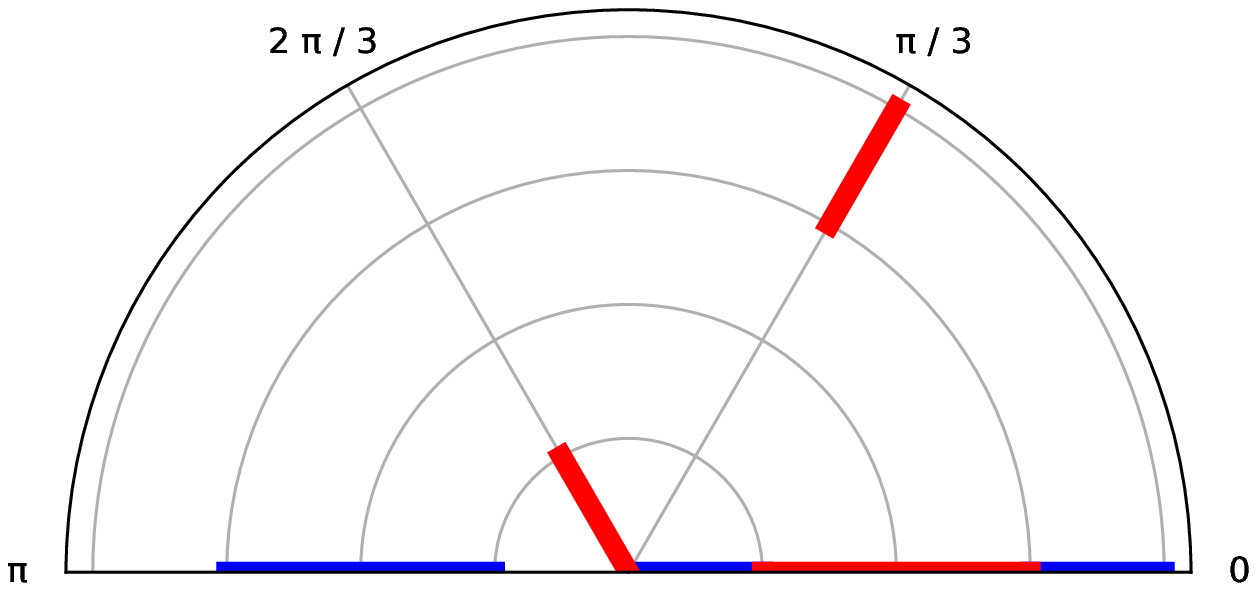}}
\caption{Force strength difference (a), and actions $(\phi, \psi)$ in (b) \boldblue{for the Red and Blue players} when following the NE for $N_r = N_b = 100$ and a tree with four actions \boldblue{for each player at each of the four decision points, with decision points indicated by the vertical green lines in panel (a)}. Parameters follow Table~\ref{Table:Parameters_1}.}
\label{fig:equilibrium_solution_100}
\end{figure*}

The influence of structural changes can be considered by setting the initial populations of both players to $100$. The equilibrium results for this scenario are presented within Figure~\ref{fig:equilibrium_solution_100}, with Red demonstrating an improving advantage across the advantage, despite several fluctuations in the penultimate stage where Blue temporarily remains stable in force strength. In the phases of panel (c), Red seeks a phase advantage in the initial and final stages with $\pi$ phase difference in the intermediate region. In this latter case, again, there appears to be periodic dynamics without the presence of any phase locking. The contrast between this outcome, and the case for $N_r = 47$ demonstrates that the game theoretic treatment introduces {\it switching dynamics} into the dynamical system of the BKL model. In some stages the Kuramoto phase dynamics may be steady state (typically when the
sought phase advantage is less than $\pi/2$), whereas others exhibit dynamical system characteristics.

\subsection{Computational Performance of Exact and Approximate Solvers}

The performance of solvers as the problem domain scales is of crucial importance. While such analysis is common for traditional numerical solutions of dynamical systems \cite{cullen2019fast, doostan2009least, nguyen2008multiscale}, the practice is less established when considering game-theoretic numerical solution concepts, with only basic examples considered in the literature \cite{matsumoto2010evaluation}. The infancy of such analysis is a product of the types of games being studied, which are either small enough that considerations of accuracy dominate concerns of computational cost (as measured in terms of calculation time); or that the games become so large that the developed solution methodologies are heavily optimised to the specific game concept, producing scaling results that are not extensible. 

\begin{figure}
\centering
  \includegraphics[width=0.7\textwidth]{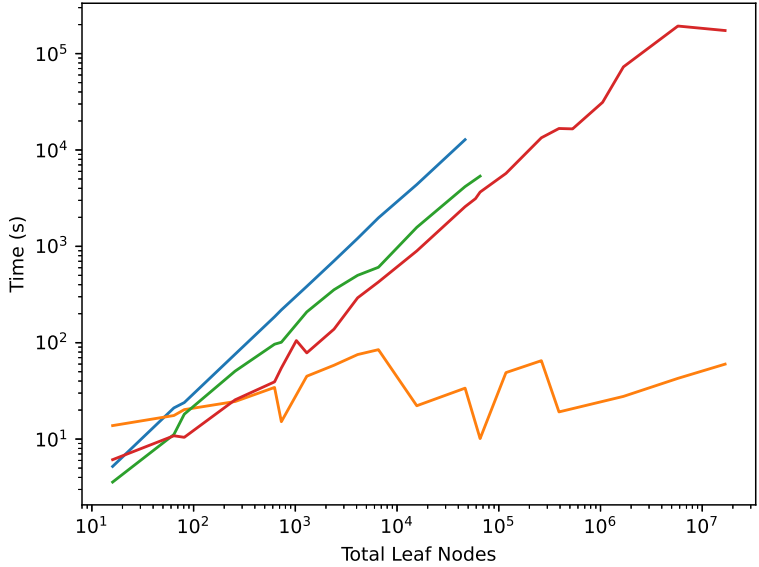}
  \caption{\boldblue{Scaling of computational cost as a function of the size of the game tree, when solving with the Full Tree (Blue), Nash Dominant (Green), MCTS performing iterations equivalent to $20\%$ of leaf nodes (Red), and Myopic (Orange) methods. Calculations are based upon the average of $6$ runs.}}
  \label{fig:LogLogScaling}
\end{figure}

\boldblue{To expand upon the extant work, the performance of the solvers of Section~\ref{sec:solutiontechniques} were tested for a range of game tree sizes, with trees defined for depths between $2$ and $4$, and action spaces for each player between $3$ and $6$. The results of this testing---as seen in Figure~\ref{fig:LogLogScaling}---demonstrate the changes in the rate of growth of computational cost as a function of the tree size. The fact that the approximate methods produce uniformly lower computational costs than the exact methods is unsurprising, however it is important to emphasise that as the size of the game tree increases, the difference between our exact Nash Dominant method and MCTS rapidly diminishes, even though MCTS is only visiting less than $20 \%$ of leaf nodes. }

\boldblue{In fact, the computational cost of the Nash Dominant solver scaled with $\mathcal{O}\left(\frac{B^{2d}}{3}\right)$ (between the theoretical upper and lower bound scalings), with the cost of MCTS exhibiting a similar scaling of $\mathcal{O}\left(\frac{2}{10} B^{2d}\right)$. It must be noted that while MCTS's ability to repeatedly visit subgame regions of a tree should produce a computational cost that is less than the number of iterations, any savings here are balanced out by the additional computational cost incurred by managing the MCTS process. The Myopic solver also conforms with the theoretical scaling properties, scaling with $\mathcal{O}\left(d B^{2}\right)$ (orange line). Based upon these results for smaller game trees, the performance advantages of the approximate solvers are not strong enough to justify their use relative to the exact Nash Dominant solver, due to approximation overheads. In larger games, Nash Dominant has the potential to notably decrease the overhead of solving the game tree relative to the Full Tree solve, while still producing an exact solution.}

Across the entire test space the Nash Dominant solver produces results that matched the exact Full Tree solutions, validating its status as an exact solver. Of the approximate solvers, Myopic produced an average error of only $2.07 \%$, with a standard deviation of $0.50 \%$, while the error from MCTS was $1.57 \%$ (with a standard deviation of $1.15 \%$). The strong relative performance of the Myopic solver across all tested tree morphologies was surprising, given that MCTS explores significantly more terminal leaves. That this is possible is a product of the structure of the BKL game itself. As the Lanchester model introduces quasi-exponential decay to the system dynamics, results at the leaves of the game tree are primarily determined by the behaviour at the initial decision points, in a fashion that favours Myopic exploration.

%%%%%%%%%%%%%%%%%%%%%%%%%%%%%%%%%%%%%%%%%%%%%%%%%%%%%%%%%%%%%%%%%%%%%%%%%%%%%%%%
\section{Parameter Analysis of Competitive Decisions to Guide Practical Applications} \label{sec:sensitive} %on Dynamical BKL Games 

Having now characterised particular regimes
of behaviour of the BKL model and
the computational performance of the Game Theory solver,
we now test the behaviour of the model across
a range of parameter values. The aim here is to
see transitions in the parameter space 
through a heatmap in Lanchester outcomes,
as originally used in \cite{Kalloniatis2020NetLanch},
from
one-side having advantage to the other side, within an
equilibrium solution and subject to
the constraints that each side brings into
the scenario (size of initial resources and their respective network C2 design, for example).
\boldblue{Treating this as a larger meta-game, the designer
of a system may detect then where risks are incurred
given their design choices.} 

Due to their analytically determined import for phase locking, the coupling parameters $\zeta_B$ and $\zeta_R$ are of particular interest, and as such we will explore $10$ distinct choices of each of these parameters. The game in question will involve $4$ actions for each of $(\phi, \psi)$ at $4$ decision points, yielding a game with $(4^2)^4=65,536$ leaf nodes. In order to better understand the nature of these games, the parameters of Table~\ref{Table:Parameters_1} are modified to decrease $\kappa_{RB}$ and $\kappa_{BR}$ to $0.002$. This change decreases the rate of attrition suffered by the resources of each through the game, to ensure that all points within the $(\zeta_B, \zeta_R )$ parameter space yield games which do not terminate early, and involve decisions being taken at all four decision points. Exploring $\zeta_B$ and $\zeta_R$ also fits the meta-game context, as they determine how tightly the agents of the player will interact within their respective decision making structures that reflects the difficulty of organisations to change how they coordinate in the midst of an adversarial engagement.

\subsection{Blue With An Initial Numerical Advantage}

\begin{figure}

\begin{minipage}{.3\linewidth}
\centering
\subfloat[Nash Dominant (Exact)]{\label{a-47}\includegraphics[scale=.35]{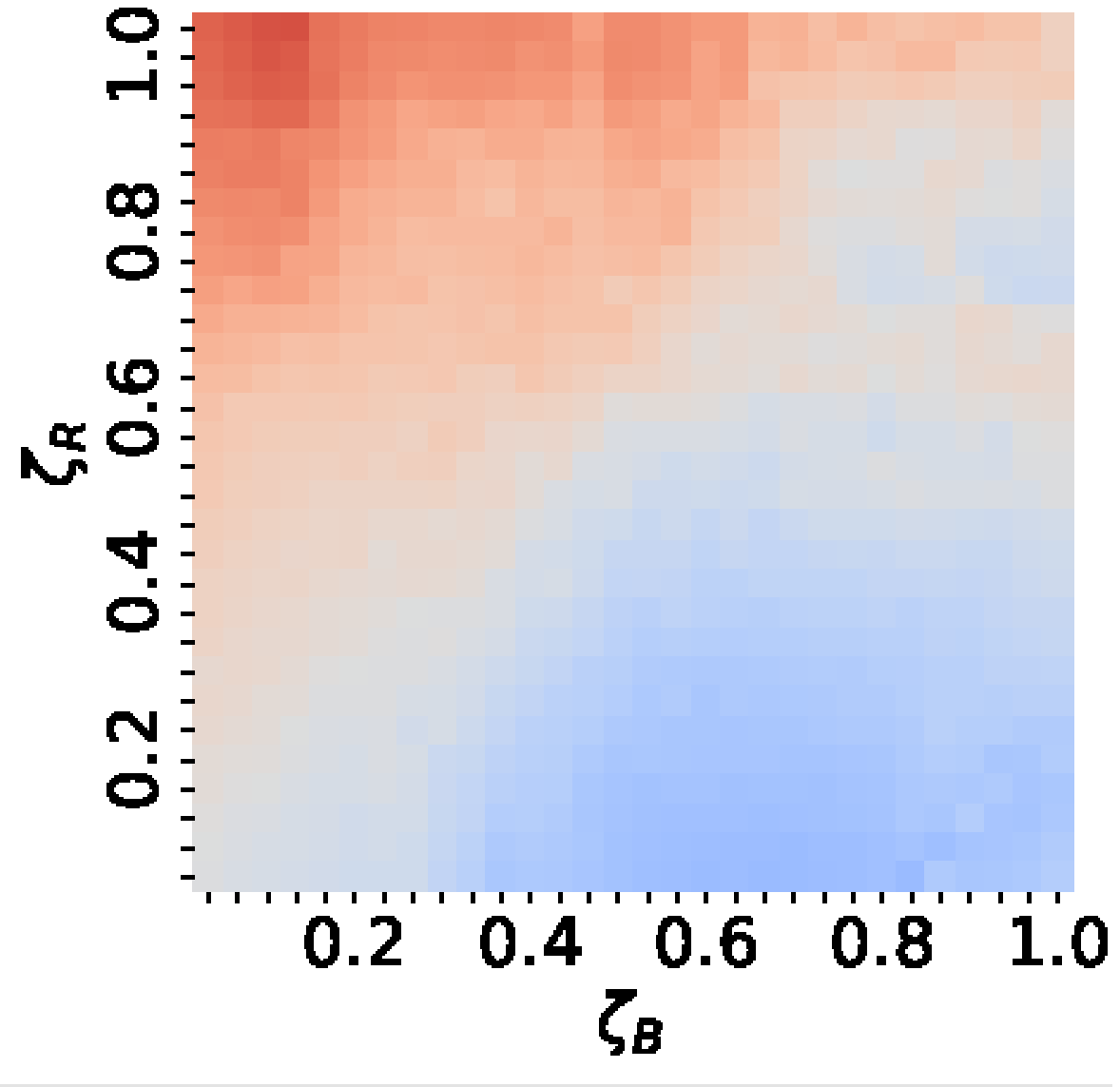}}
\end{minipage}%
\begin{minipage}{.3\linewidth}
\centering
\subfloat[Myopic (Approx.)]{\label{b-47}\includegraphics[scale=.35]{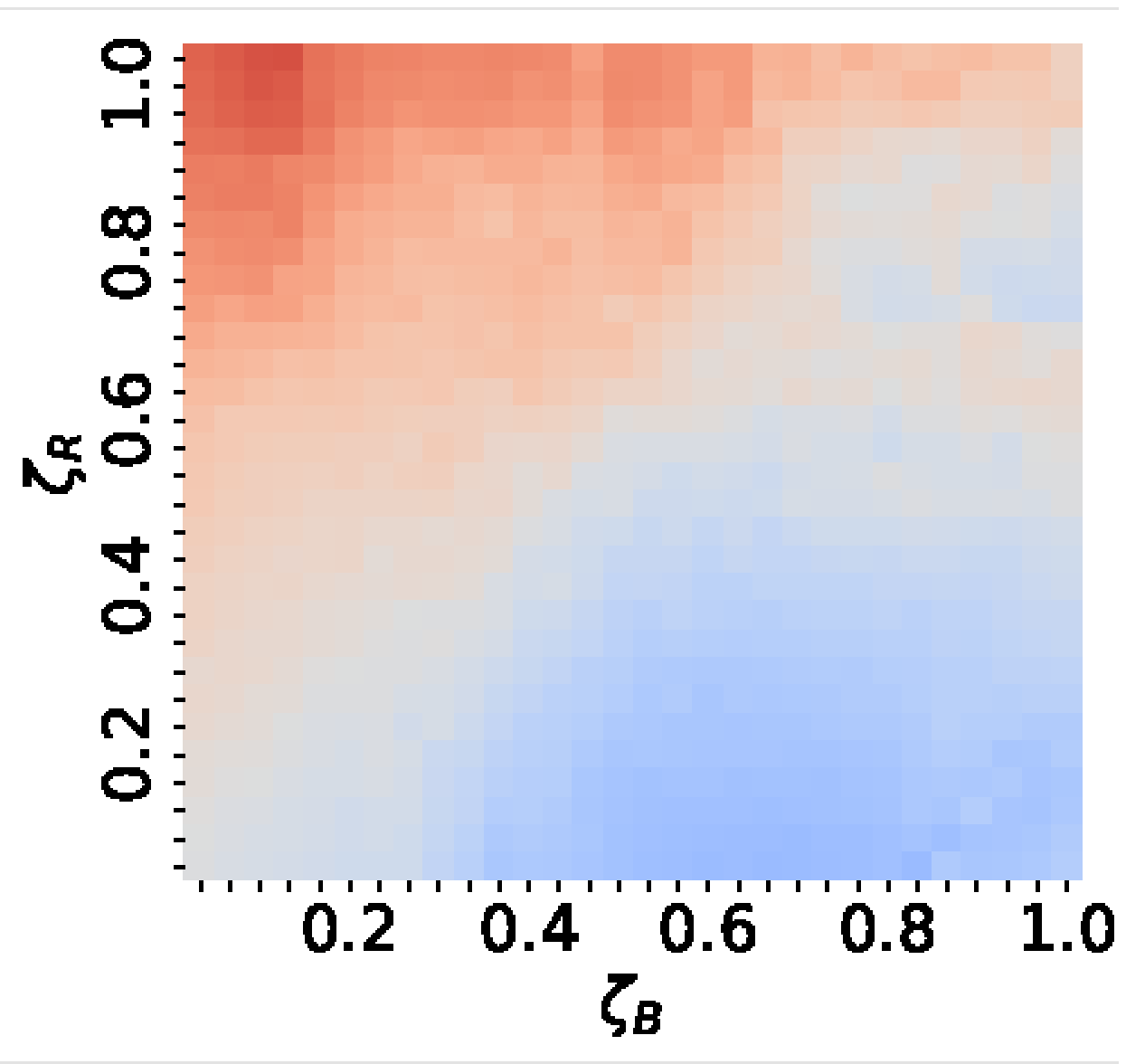}}
\end{minipage}
\begin{minipage}{.3\linewidth}
\centering
\subfloat[MCTS (Approx.)]{\label{c-47}\includegraphics[scale=.35]{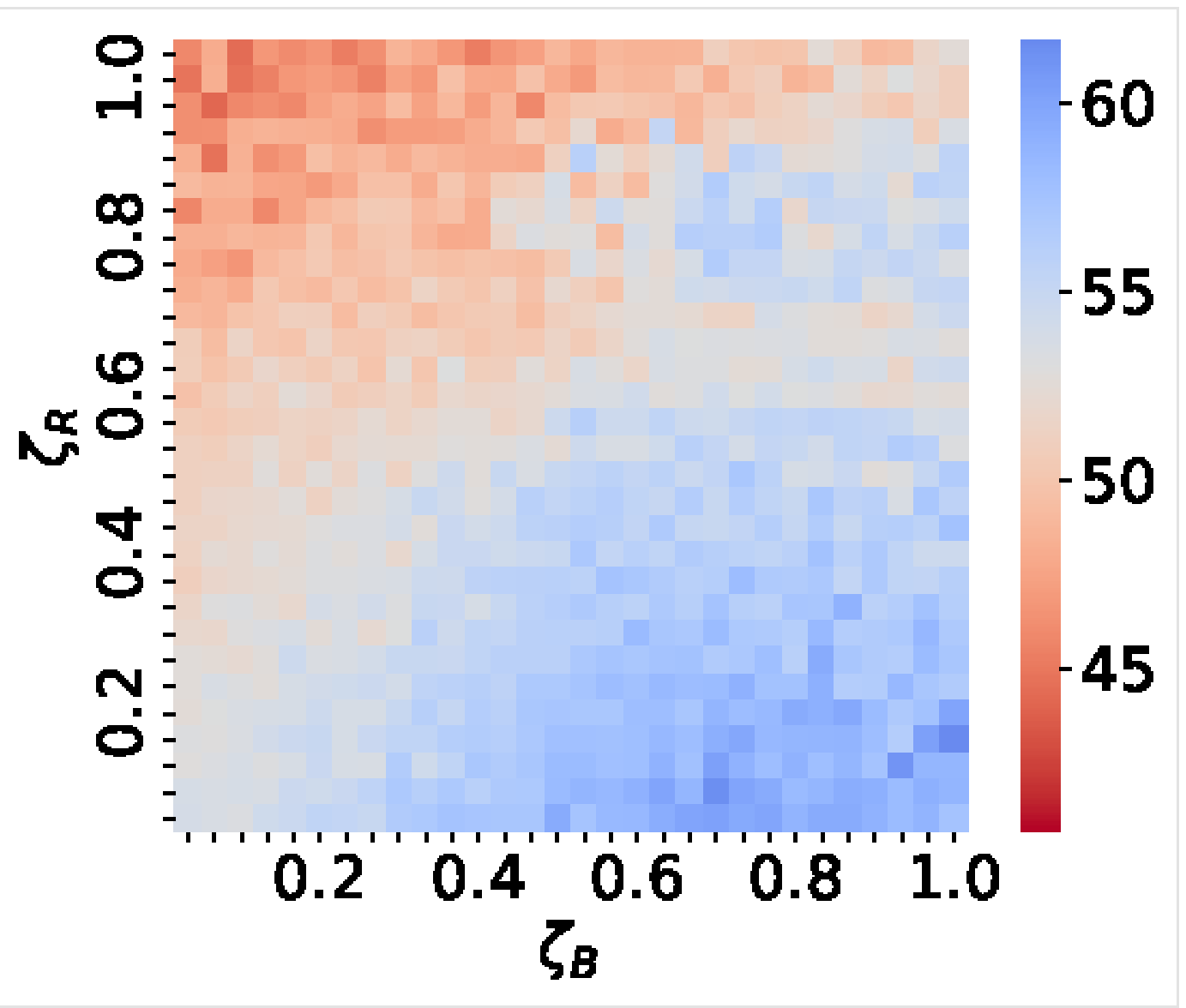}}
\end{minipage}
\caption{\boldblue{Exploring the $(\zeta_B, \zeta_R)$ solution spaces across the solvers for $N_r = 47.$ Here red colours denote states more favourable to the Red player than the initial population difference, with the same for blue colours and the opponent.}}
\label{fig:Consolidated-47}
\end{figure}

For the case where the initial resources are $(N_r, N_b) = (47, 100)$,  Figure~\ref{fig:Consolidated-47} explores the constructed solution space as per the Nash Dominant, Myopic and MCTS solvers. We emphasise that a red colour does not indicate
defeat for Blue for which a negative value of its utility would be required, but rather the colour reflects the degree attrition; and that the heat maps correspond not to fixed $(\phi, \psi)$, but rather the equilibrium actions at each $(\zeta_B, \zeta_R)$ pairing. Across all solution concepts, the evolution of the game state provides a broad preservation of this advantage across the tested range of game states, with an overall range of admitted equilibrium utilities in $[42.5, 60.3]$. The player who gains advantage over the course of the game, by either increasing (for Blue) or decreasing (for Red) the overall utility relative to the initial equilibrium state of $53$, is determined by the relative balance of $\zeta_B$ and $\zeta_R$. 

At face value the plot is consistent with the basic intuition that
stronger relative internal coupling is beneficial. Thus, when $\zeta_B < \zeta_R$, the admitted NE state favours Red, with the utility reaching a peak when $(\zeta_B, \zeta_R) = (0.05, 1)$. Where $\zeta_B > \zeta_R$ Blue is almost uniformly favoured, \boldblue{with the exception of a small band of isolated Red favoured results at $\zeta_R \approx 0.6$, and a general trend towards Red favourable outcomes as $\zeta_B \to 1$}. While there is a monotonic increase in utility for Red with respect to increases in $\zeta_R$ over the range considered, the same cannot be said for the Blue response to $\zeta_B$. Instead  the numerically superior Blue player exhibits a `sweet spot' at $\zeta_B \approx 0.7$, with further increases to $\zeta_B$ producing diminishing results across all choices of $\zeta_R$. This is consistent \boldblue{with previous work} \cite{Ahern2020}, as excessive internal coupling in relation to the internal structure and other variables
leads to a `rigidity' in the system. A similar point of diminished returns
exists also for Red beyond the range of $\zeta_R$
considered here, \boldblue{where the broader range} is a consequence of the higher connectivity of Red compared to the hierarchical structure of Blue.

While these observed behaviours are driven by the players attempting to find an equilibrium solution, the dynamics are still tied to the underlying dynamical equations. As per Equation~\ref{e:kuramoto1}, increasing $\zeta_B$ (relative to \boldblue{$\zeta_{BR}$}, which is fixed at $0.4$ for these experiments) increases the importance of spreads in $\beta$ to the overall derivative $\frac{d \beta}{dt}$, and thus, in Blue's positioning against Red. Increasing this component of Equation~\ref{e:kuramoto1} allows for the Blue player to more precisely tune its own evolution, in a fashion that is only weakly coupled to the behaviour of Red, allowing the Blue player to theoretically eke out greater advantages in terms of its organisational positioning, and, in turn, the overall utility of the game. There is a symmetrical behaviour in Red, however the differences in the underlying force organisations---hierarchical for Blue, and unstructured for Red---dictate the asymmetry in the observed responses to changing $(\zeta_B, \zeta_R)$. Considering changes in these parameters as part of a larger meta-game in turn gives that the meta-game itself must also have an equilibrium state. In this example, this occurs at $(\zeta_B, \zeta_R) = (1, 1)$, yielding an overall NE utility of $52$. This corresponds to Blue retaining a numerical advantage over the time period of the engagement.

Both the Myopic (b) and MCTS (c) solvers accurately capture the dynamics exhibited by the exact solution (a), with errors consistent with Section~\ref{sec:NumericalAnalysis}. While under visual inspection both approximate solvers broadly replicate the dynamics of the exact solution, the MCTS solution exhibits the smallest absolute error, with a mean, max, and standard deviation of the absolute errors of $(0.465, 0.530,  2.369)$, as compared to $(0.635, 0.458, 2.372)$ from the Myopic solver. The differences between the solution methodologies can primarily be seen at the extremum of $\zeta_B$ and $\zeta_R$. The Myopic solver consistently overestimates the values in the regions where the equilibrium is at its largest, although it does accurately capture the location of the best response solutions for each player (the location of the largest Red and Blue favoured scores). In contrast, while the MCTS solver is slightly more accurate overall, it fails to confidently capture the best response solutions for each player, although it does still capture the equilibrium of the meta-game.

The correspondence between the Nash Dominant and Myopic solutions is due to the tested position in parameter space is heavily dominated by earlier decisions in the parameter space. This is a consequence of the Lanchester models quasi-exponential resource decay, with the end-state Nash Equilibrium dominated by the decisions at the earliest game states. The Myopic solver also outperforms as the action space for each player minimally changes as they move deeper into the game tree. As such, there's no incentive for players to make sub--optimal decisions in the early game states---which heavily influence the overall evolution of the equilibrium---in order to open up parts of the game tree that are more favourable as the game progresses. We hypothesise that extending the game to one where actions at one decision point influence the available action space at the next decision point would lead to the Myopic solver to under perform relative to MCTS.

Considering $N_r=80$ reveals broad structural similarities to the prior case. The primary change is that the final utilities are uniformly lower. Notably, the region of $(\zeta_B, \zeta_R)$ space where Blue improves upon their starting position (relative to Red) has decreased in both area and peak magnitude. The location of this peak magnitude, the point of best response for Blue, has also shifted slightly to the left, to weaker coupling, 
compared to Fig.\ref{fig:Consolidated-47} although discerning the exact nature of the shift is complicated by the discretisation.
\boldblue{For brevity, we do not show a plot of this case.} 

\begin{figure}

\begin{minipage}{.3\linewidth}
\centering
\subfloat[Nash Dominant (Exact)]{\label{a-100}\includegraphics[scale=.35]{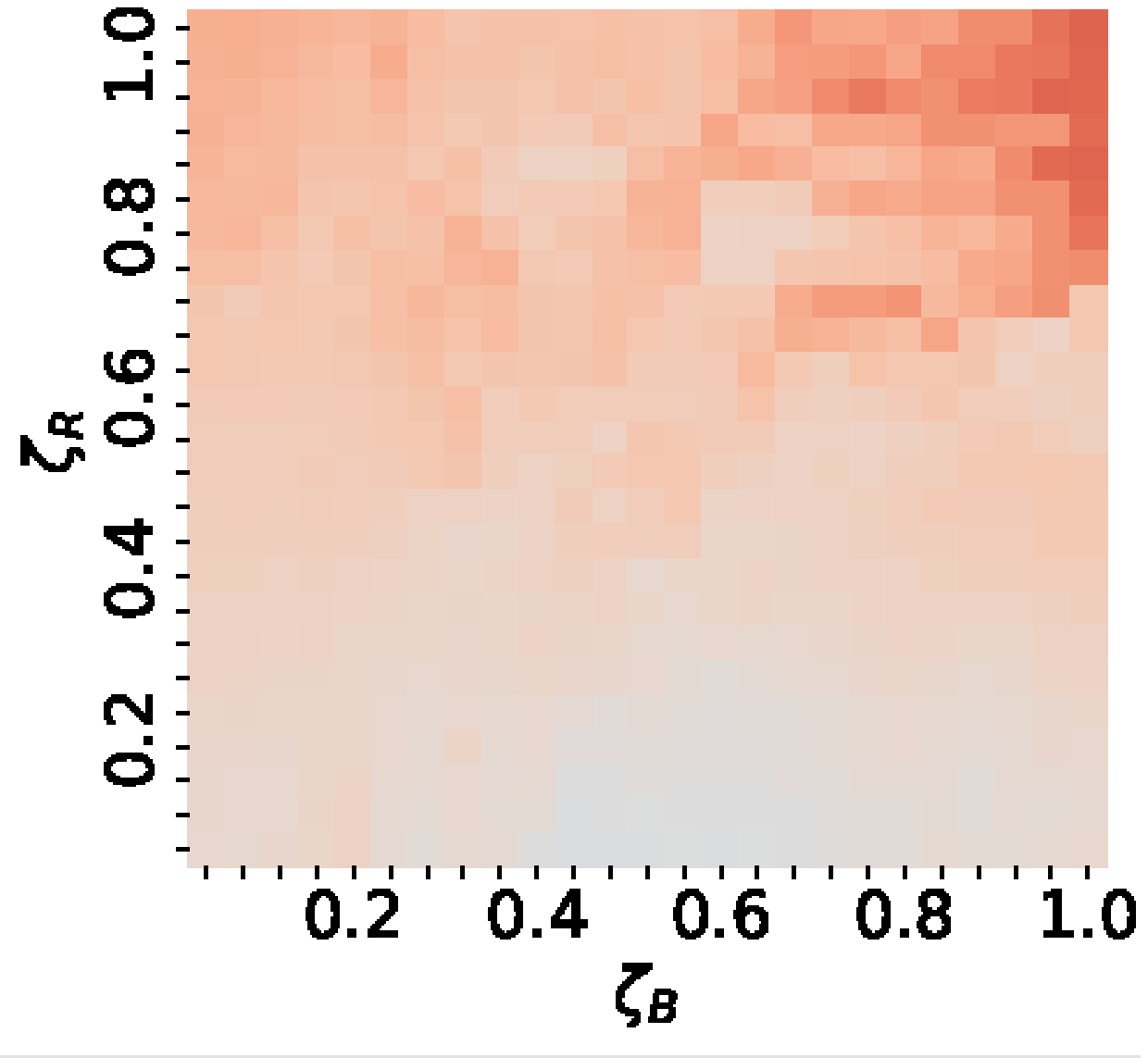}}
\end{minipage}%
\begin{minipage}{.3\linewidth}
\centering
\subfloat[Myopic (Approx.)]{\label{b-100}\includegraphics[scale=.35]{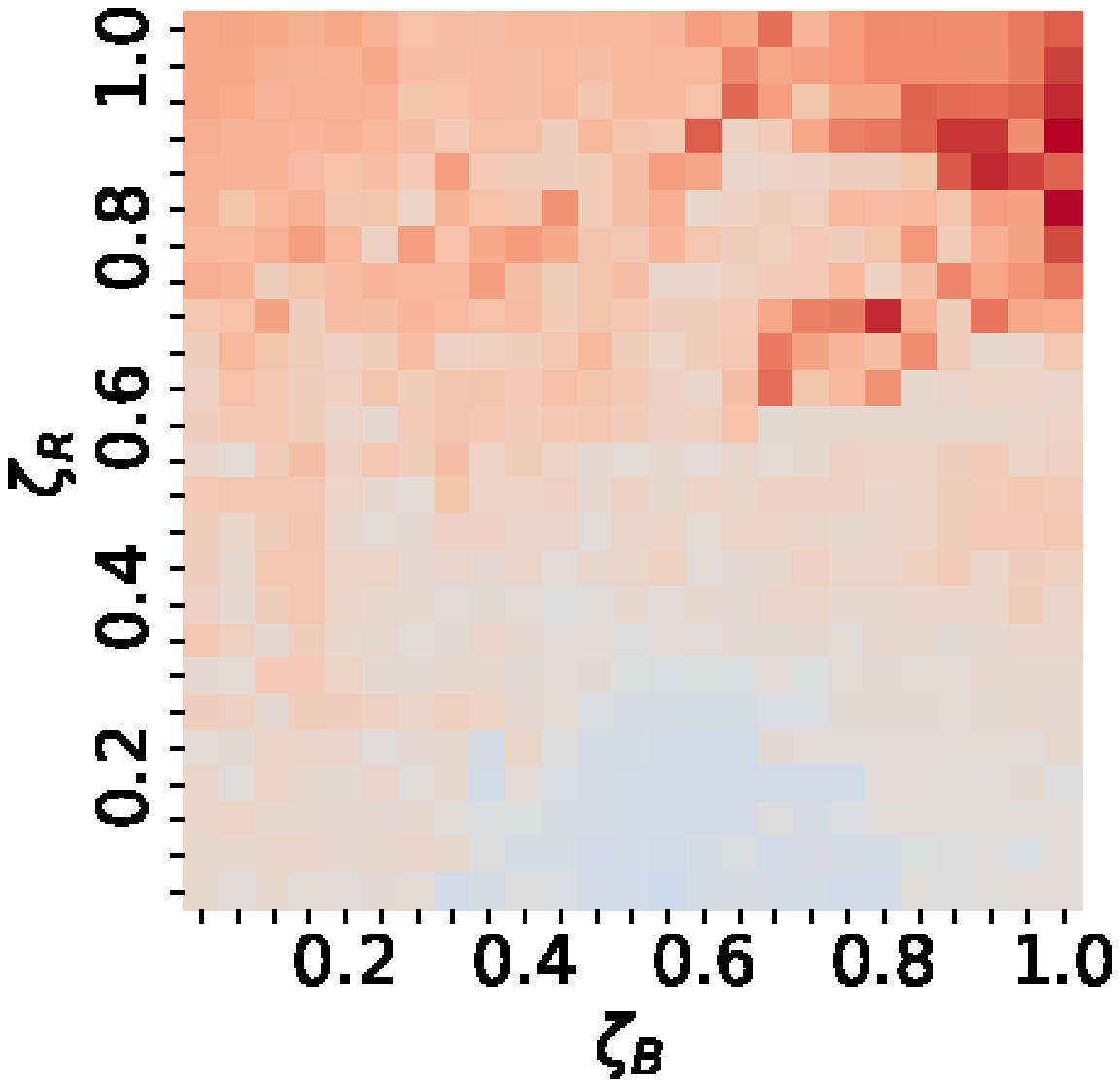}}
\end{minipage}
\begin{minipage}{.3\linewidth}
\centering
\subfloat[MCTS (Approx.)]{\label{c-100}\includegraphics[scale=.35]{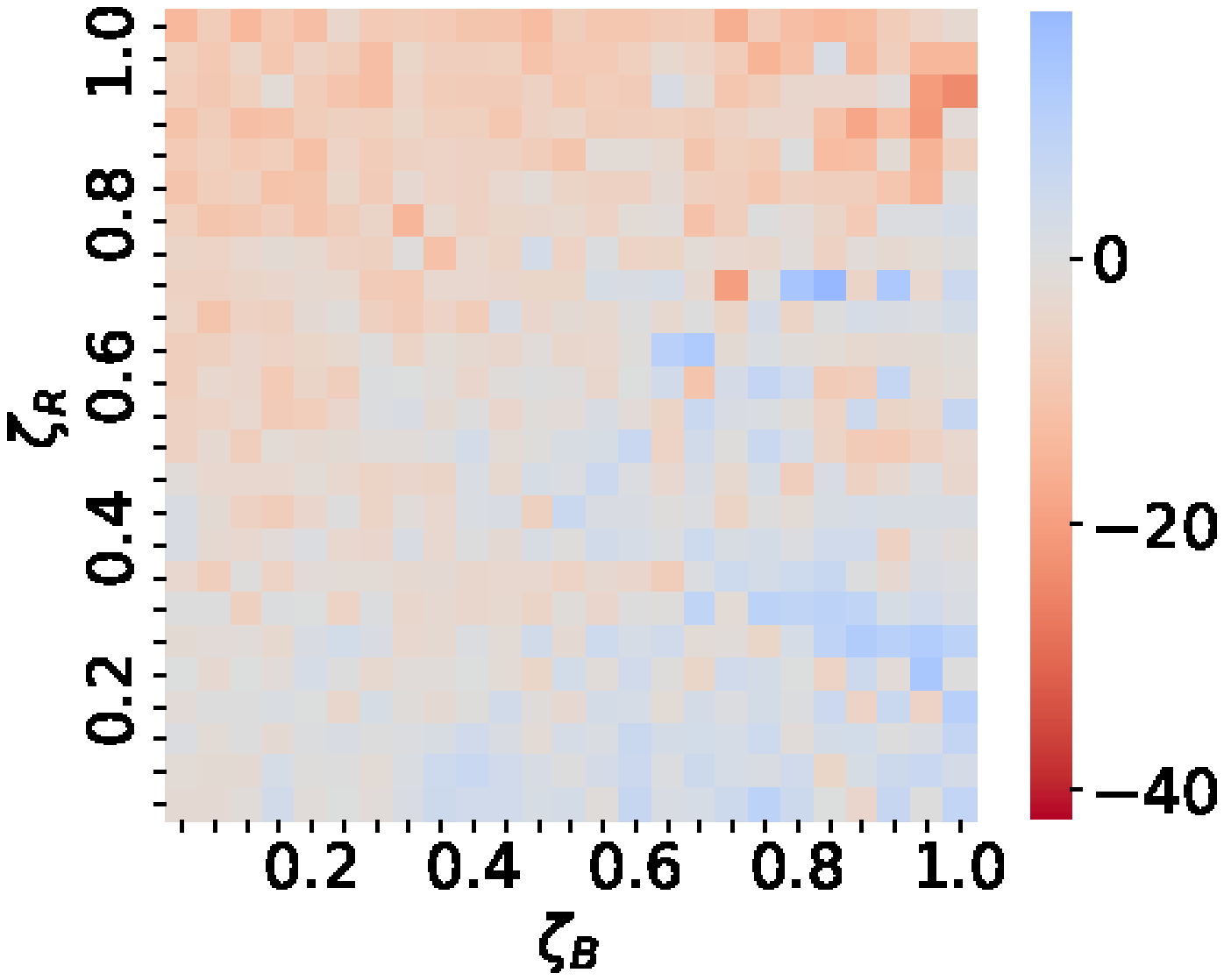}}
\end{minipage}
\caption{\boldblue{Exploring the $(\zeta_B, \zeta_R)$ solution spaces across the solvers for $N_r = 100.$ Here red colours denote states more favourable to the Red player than the initial population difference, with the same for blue colours and the opponent.}}
\label{fig:Consolidated-100}
\end{figure}

\subsection{Blue and Red Under Initial Parity}
\boldblue{In a parity situation where both populations are initially} $100$,  Figure~\ref{fig:Consolidated-100} shows that instead of the clear distinction between blue-dominant and red-dominant regions in response to the $(\zeta_B, \zeta_R)$ space, the Nash Equilibria without regions of local homogeneity, in which small changes in $\zeta_B$ or $\zeta_R$ can lead to significant differences in the utility. While increasing $\zeta_R$ still does, in general, lead to increases in utility for the Red player in this case, there are small isolated regions as $\zeta_R \to 1$ which \boldblue{are more favourable to} the Blue player, relative to the surrounding positions in parameter space. The presence of these isolated regions is driven by the individual solution profiles \boldblue{in the parity case} exhibiting more chaotic solution dynamics, and a broader range of utilities within an individual game. This also drives the greater range in equilibrium solutions exhibited within Figure~\ref{fig:Consolidated-100}. %is also driven by this. 

While it would be expected that further increases to $N_r$ would increase the proportion of the $(\zeta_B, \zeta_R)$ domain in which Red is favoured, the observed changes to the equilibrium outcomes are not uniform. Notably, the `sweet spot' for Blue is now lower in coupling value and weaker in utility: through increases in Red's initial force over the values $N_r=\{47, 80, 100\}$, the maximal point of utility for Blue has steadily decreased, spanning values of $\zeta_B = \{0.7, 0.58, 0.4\}$. That these systems exhibit sensitivity to the initial conditions is a hallmark of the underpinning chaotic dynamical system. This equal initial resource parity scenario creates the greatest sensitivity to other parameters where the decision making process---as reflected through the phase dynamics---plays the dominant role. This is reflected in the time-dependent view of Figure~\ref{fig:equilibrium_solution_100}, which corresponds to the centre of the heatmap in Figure~\ref{fig:Consolidated-100}, where Red's phase advantage is the factor that leads to its superior
resource strength at the end of the dynamics. 

These dynamics demonstrate why the equal coupling $\zeta_B=\zeta_R$ for  Figure~\ref{fig:equilibrium_solution_100} leads to Blue defeat---most of the heatmap leads to such outcomes as a consequence of Blue's less favourable network structure. That the optimal response%, in terms of the coupling parameters,
can be influenced by the initial relative force strengths accords with known domains in which there is interplay between competitive organisations. As alluded earlier, Blue is persistently at a disadvantage through its hierarchical structure, gaining advantage only through initial superior resource and applying tighter internal effort in its decision making.

\subsection{Analysis of Solution Concepts}

To quantitatively assess the overall performance of the approximate solvers relative to the exact Nash Dominant Solutions, Table~\ref{tab:normalised_results} considers the average absolute error, the normalised average absolute error---when normalised against the range of solutions seen over $(\zeta_B, \zeta_R)$ and standard deviation of error across each of the tested scenarios. While the Myopic solution concept under-performs relative to MCTS when $N_r = 47$, increasing $N_r$ yields a notable deterioration in the performance of the more computationally expensive MCTS solution concept. Even when the number of iterations is increased to $45\%$ of the total number of leaf nodes, the MCTS average absolute error only decreases by $13.5 \%$, producing results that are still inferior to the Myopic solver.

\begin{table}[h]
\centering
\scalebox{0.7}{
\begin{tabular}{|c|c|c|c|c|c|}
\hline
$N_r$   & Solver & Mean & Normalised Mean & Std & Norm. Std \\
\hline
$47$ & MCTS & $0.465$ & $0.027$ & $0.530$ & $0.031$ \\
     & Myopic & $0.635$ & $0.037$ & $0.458$ & $0.027$ \\
     \hline
$80$ & MCTS & $1.341$ & $0.056$ & $1.137$ & $0.055$ \\
     & Myopic & $0.885$ & $0.038$ & $0.777$ & $0.038$ \\
     \hline
$100$ & MCTS & $6.440$ & $0.192$ & $3.558$ & $0.106$ \\
      & Myopic & $4.469$ & $0.133$ & $3.247$ & $0.097$\\
\hline
\end{tabular}
}
\caption{Mean and Standard Deviation of the absolute errors for
final Blue utility values, and their normalised equivalents for all tested cases. Normalisation performed by dividing by the range of equilibrium states admitted across all solutions for $N_r = 47$ and $100$ respectively.}
\label{tab:normalised_results}
\end{table}

 The divergence between the $N_r = 47$ and $80$ cases---both of which have reasonably well structured meta-game spaces that show monotonic changes---and the more variable $N_r=100$ case is stark, with the latter scenario exhibiting errors that are up to $1$ order of magnitude larger than the equivalent $N_r = 47$ solutions. While the absolute errors for both solvers are approximately equivalent, the Myopic solutions err by overestimating both the most Red and Blue favoured equilibrium states; and MCTS biases the solutions towards the weaker player. That this occurs indicates that even with the corrections we have made to the MCTS algorithm, it is still more suited to games that approach binary win-loss states, and struggles to accurately resolving solutions in the Red favoured $N_r=100$ game-state.

\section{Conclusion}
We have shown that physics-based dynamical 
models may be employed to model competitive decision-making processes. Such a model is made possible through the incorporation of Game Theory, and yields insights that are both intuitively reasonable and have real world applications. 
 
When exploring the parameter space of a domain of industrial significance, we uncovered that a player with greater internal connectivity was able to more appropriately position its response to an opposing player. This was true even in the face of a significant numerical disadvantage. For the hierarchically structured player we observed a limit in how far increasing internal coupling improved their position: beyond a certain point increases became counter-productive with unfavourable results against a more agile connected counterpart. This demonstrates that coupling is not simply interchangeable with network connectivity: increased connectivity increases the range of coupling strength over which advantage can be gained over a less connected competitor.

This work also uncovered that as the game moves away from a balanced equilibrium, the outcome of the game transitions away from a smooth response under changes in the parameters. The existence of such behaviours underscores the importance of being able to accurately solve these games in a numerically efficient manner, in order for players to most advantageously position themselves in competition.

In aide of this, we developed a novel exact solver, and tested several established approximate numerical schemes. While the approximate solvers were able to accurately resolve the dynamics when the game was in a balanced state, they failed to accurately resolve imbalanced states. These issues are pronounced for MCTS, which is more suited for win--loss games rather than those with continuous outputs. In contrast to the behaviours of the established approximate solvers, our new Nash Dominant numerical solver was able to efficiently construct exact numerical solutions to these games in a numerically efficient fashion. 
Future work will apply this solver to the more
computationally demanding form of the networked
BKL model in \cite{Ahern2020}.

\section*{Acknowledgment}

This research was funded in part by the Commonwealth of Australia through the Modelling Complex Warfighting Strategic Research Initiative of the Defence Science and Technology Group. Dr. Andrew Cullen was with the Department of Electrical and Electronic Engineering, The University of Melbourne during part of this research.

\bibliographystyle{IEEEtran}
\bibliography{Arxiv}

\end{document}